\documentclass[aip,pop,preprint,supersecriptaddress,showpacs]{revtex4-1}

\usepackage[centertags,reqno]{amsmath}
\usepackage{amssymb}
\usepackage{graphicx} 
\usepackage{epsfig}
\usepackage{ulem}
\usepackage{eso-pic}                                                            
\def\beq{\begin{equation}}
\def\eeq{\end{equation}}
\def\beqar{\begin{eqnarray}}
\def\eeqar{\end{eqnarray}}
\def\nn{\nonumber}

\def\para{\parallel}
\def\etal{{\itshape et al.}}

\newcommand{\pdiff}[2]{\frac{\partial#1}{\partial#2}}

\newcommand{\pdt}{\partial_t}
\newcommand{\pdr}{\partial_r}
\newcommand{\pdth}{\partial_\theta}

\newcommand{\enum}[2]{{#1}\times10^{#2}} % 4.2x10^{3} = \enum{4.2}{3}

\def\div{\nabla\cdot}
\def\grad{\nabla}

\newcommand{\gradpar}{\grad_\parallel}
\newcommand{\gradperp}{\grad_\perp}

\newcommand{\defeq}{\ensuremath{\stackrel{\text{\tiny def}}{=}}}

\newcommand{\savg}[1]{\left<{#1}\right>}
\newcommand{\vavg}[1]{\left<{#1}\right>_V}
\newcommand{\thavg}[1]{\left<{#1}\right>_\theta}

\newcommand{\vpar} {v_\parallel}

\newcommand{\vE}{\ensuremath{\boldsymbol{{\rm v}_E}}}
\newcommand{\bo}{\ensuremath{\boldsymbol{{\rm b}_0}}}
\newcommand{\bvec}{\ensuremath{\boldsymbol{{\rm b}}}}
\newcommand{\xvec}{\ensuremath{\boldsymbol{{\rm x}}}}

\newcommand{\zvec}{\ensuremath{\boldsymbol{{\rm z}}}}
\newcommand{\vvec}{\ensuremath{\boldsymbol{{\rm v}}}}
\newcommand{\jvec}{\ensuremath{\boldsymbol{{\rm j}}}}

\newcommand{\vpe}{v_{\parallel e}}

\newcommand{\nue}{\nu_{e}}

\newcommand{\nuin}{\nu_{in}}

\newcommand{\wci}{\Omega_{ci}}
\newcommand{\wcix}{\Omega_{cix}}

\newcommand{\Isat}{I_{i,sat}}

\newcommand{\cm}{\rm cm}

\newcommand{\cmn}{{\rm cm}^{-3}}
\newcommand{\mn}{{\rm m}^{-3}}
\newcommand{\eV}{\rm eV}
\newcommand{\G}{\rm G}
\newcommand{\T}{\rm T}

\begin{document}

\title{Modeling of plasma turbulence and transport in the Large Plasma
Device}
\author{P. Popovich}
\affiliation{Department of Physics and Astronomy and Center for
  Multiscale Plasma Dynamics, University of
  California, Los Angeles, CA 90095-1547}
\author{M.V. Umansky}
\affiliation{Lawrence Livermore National Laboratory,
Livermore, CA 94550, USA}
\author{T.A. Carter}
\email{tcarter@physics.ucla.edu}
\author{B. Friedman}
\affiliation{Department of Physics and Astronomy and Center for
  Multiscale Plasma Dynamics, University of
  California, Los Angeles, CA 90095-1547}


%52.30.Ex 	Two-fluid and multi-fluid plasmas 
%52.35.Fp 	Electrostatic waves and oscillations (e.g., ion-acoustic waves)
%52.35.Kt 	Drift waves
%52.65.Kj 	Magnetohydrodynamic and fluid equation
%52.35.Qz 	Microinstabilities (ion-acoustic, two-stream, loss-cone, beam-plasma, drift, ion- or electron-cyclotron, etc.)
%52.35.Ra 	Plasma turbulence 
%52.25.Fi 	Transport properties 

\date{\today}

\begin{abstract}

Numerical simulation of plasma turbulence in the Large Plasma Device
(LAPD) [Gekelman \etal, Rev. Sci. Inst., 62, 2875, 1991] is
presented. The model, implemented in the BOUndary Turbulence (BOUT)
code [M. Umansky \etal, Contrib. Plasma Phys. 180, 887 (2009)],
includes 3-D collisional fluid equations for plasma density, electron
parallel momentum, and current continuity, and also includes the
effects of ion-neutral collisions. In nonlinear simulations using
measured LAPD density profiles but assuming constant temperature
profile for simplicity, self-consistent evolution of
instabilities and nonlinearly-generated zonal flows results in a
saturated turbulent state.  Comparisons of these simulations with
measurements in LAPD plasmas reveal good qualitative and reasonable
quantitative agreement, in particular in frequency spectrum, spatial
correlation and amplitude probability distribution function of density
fluctuations. For comparison with LAPD measurements, the plasma
density profile in simulations is maintained either by direct
azimuthal averaging on each time step, or by adding particle
source/sink function. The inferred source/sink values are consistent
with the estimated ionization source and parallel losses in LAPD.
These simulations lay the groundwork for more a comprehensive
effort to test fluid turbulence simulation against LAPD data.

\end{abstract}

\maketitle

%%%%%%%%%%%%%%%%%%%%%%%%%%%%%%%%%%%%%%%%%%%%%%%%%%%%%%%%%%%%%%%%%%%%%%%%%%%

\section{Introduction}

Turbulent transport of heat, particles and momentum has an impact on a
wide variety of plasma phenomena~\cite{schekochihin2006,
  quataert2002,hasegawa2004}, but is of particular importance for
laboratory magnetic confinement experiments for fusion energy
applications~\cite{tynan2009,connor1995,horton1999, zweben2007}.  A large number of advances in understanding
plasma turbulence have been made using analytic theory, for example
nonlinear mode interaction~\cite{Kadomtsev1965,Stenflo1970},
instability saturation and secondary
instabilities~\cite{cowley1991,dorland2000},
cascades~\cite{Iroshnikov1963,Kraichnan1965} and the role of sheared
flow~\cite{Itoh1988,Shaing1989,Biglari1990}.  However it is
increasingly necessary to use direct numerical simulation as a tool to
gain understanding into the complex nonlinear problem of plasma
turbulence.  Additionally, numerical simulation is key to the
development of a predictive capability for turbulent transport in
fusion plasmas.  An essential aspect of the development of this
capability is its validation of numerical simulation against
experimental measurement~\cite{Greenwald2010,Terry2008}.

While ultimately validation against measurements in high-temperature
fusion plasmas in toroidal geometry must be undertaken, it is
desirable to have a hierarchy of experiments for comparison, with the
goal of isolating important physical effects in the simplest possible
geometry~\cite{Greenwald2010}.  Linear plasma devices such as
LAPD~\cite{Gekelman1991}, CSDX~\cite{Tynan2004},
VINETA~\cite{Franck2002}, LMD~\cite{Shinohara1995},
HELCAT~\cite{Lynn2009}, and MIRABELLE~\cite{Pierre1987} offer an
opportunity to validate turbulence and transport simulations in simple
geometry and with boundary conditions and plasma parameters with
reasonable relevance to tokamak edge and scrape-off-layer plasmas.
Thanks to their low temperature, these
devices are highly diagnosable, providing for detailed comparison
against code predictions.  As these plasmas tend to be fairly
collisional, fluid (including gyrofluid) models have been compared in
recent studies, for example, on LMD~\cite{kasuya2007},
CSDX~\cite{Holland2007}, and
VINETA~\cite{Naulin2008,Kervalishivili2008,Windisch2008}.  These
studies have not simply compared code output to data, but more importantly
have been used to extract physics understanding: the importance of
ion-neutral collisions in zonal flow damping was explored in
LMD~\cite{kasuya2007}; simulations of the VINETA device were focused
on exploring the formation and propagation of turbulent
structures~\cite{Naulin2008,Kervalishivili2008,Windisch2008}; and
recent simulations of the LAPD plasma suggest that sheath boundary
conditions in some regimes could drive strong potential gradients and
in this case the Kelvin-Helmholtz instability can dominate over drift-type
instabilities~\cite{Rogers2010}.  

This paper presents modeling of turbulence and transport in the Large
Plasma Device (LAPD)~\cite{Gekelman1991} using a Braginskii fluid
model implemented in the BOUndary Turbulence (BOUT)
code~\cite{Xu1998,Umansky2009}.  LAPD provides a unique platform for studying
turbulence and transport.  Large size perpendicular to the magnetic field ($100 \lesssim
\rho_i/a \lesssim 300$) results in a large number of linearly unstable
modes~\cite{Popovich2010a} and broadband, fully developed turbulence is
  observed~\cite{carter2009}.  Due to its length (17m), perpendicular
transport can dominate over parallel losses and changes in turbulent
transport can have a strong impact on radial plasma
profiles~\cite{maggs2007}.  The LAPD plasma is similar to tokamak SOL
in the sense that the radial plasma density and temperature profiles
are determined by the competition of the radial turbulent transport,
parallel streaming, and volumetric sources.  The use of the BOUT code
also provides a unique opportunity to directly test in linear geometry 
the same code that is also used to simulate tokamak edge plasmas

Numerical simulations reported here are done in LAPD geometry using
experimentally measured density profiles.  In order to simplify
these initial studies, a flat temperature profile is assumed, flow
profiles are allowed to freely evolve in the simulation, and periodic
axial boundary conditions are employed.  The simulations show a
self-consistent evolution of turbulence and self-generated electric
field and zonal flows. The density source/sink required to maintain
the average density profile close to the experimental profile is
consistent with the ionization source and parallel streaming losses
in LAPD. Overall, these calculations appear to give a good qualitative
and reasonable quantitative match to  experimental
temporal spectra and are also consistent with the measured spatial
structure and distribution of fluctuation amplitude.  These
results lay the foundation for proceeding with the more difficult
task of simulations with matched density, temperature and flow
profiles along with more realistic axial boundary conditions.

This paper is organized as follows. In Section~\ref{secModel} the main
parameters of LAPD are described, as well as the fluid equations
implemented in the BOUT code that are used to model the LAPD device.
Section~\ref{secAvgProfile} discusses the two methods of average
profile control that are used to maintain the average density close to
the experimental values. A detailed comparison of simulated turbulence
characteristics to the experimentally measured quantities is presented
in Section~\ref{secCompare}. Section~\ref{secDiscussion} discusses the
particle transport and the density source in the simulations, and also
briefly introduces the numerical diagnostics used for verifying the
solution. Conclusions are presented in Section~\ref{secConclusions}.
The appendices demonstrate the derivation of the azimuthal momentum
equation (Appendix~\ref{secAppendixEr}), derivation and discussion of
the ion viscosity term (Appendix~\ref{secAppendixVisc}) and a
numerical scheme used to avoid unphysical solutions due to parallel
discretization (Appendix~\ref{secAppendixGrid}).

%%%%%%%%%%%%%%%%%%%%%%%%%%%%%%%%%%%%%%%%%%%%%%%%%%%%%%%%%%%%%%%%%%%%%%%%%%%

\section{Physics model}\label{secModel}

The LAPD device is a long cylindrical plasma configuration
with length $L\sim 17$~m, vacuum vessel radius $r_s$=50 cm, typical plasma
radius $a\sim 30~\cm$, electron density $n_{e0}
\lesssim 5\times10^{12}~\cmn$, electron temperature $T_e\lesssim 10~\eV$,
and ion temperature $T_i\lesssim 1~\eV$; with an externally imposed
axial magnetic field magnetic field $B_z<0.25~\T$. Plasmas are
typically composed of singly ionized helium although argon, neon and
hydrogen plasmas can also be studied.

For the calculations discussed here, LAPD is modeled as a cylindrical
annulus with inner radius 15 cm and outer radius 45 cm. Using the
annulus topology allows the LAPD geometry to be described in the BOUT
code without any modification of the code itself through using the
built-in tokamak geometry but changing the metric coefficient
values~\cite{Popovich2010a}. In this setup, the poloidal magnetic
field of the tokamak configuration corresponds to the axial field of
LAPD, and the toroidal field is set to zero as it corresponds to
the azimuthal direction in LAPD. The annulus configuration also avoids
the potential complications of the cylindrical axis singularity.
The magnetic field is taken uniform, along the cylinder axis, and
the axial boundary conditions are taken to be periodic. 

The simulations presented here are based on the Braginskii two-fluid
model~\cite{Braginskii1965}. As discussed in a linear verification
study~\cite{Popovich2010a} that uses the same model, collisions play a
very important role in LAPD plasmas. Electron-ion collision rate is
much higher than the characteristic drift frequencies, $\nu_{ei} \gg
\omega_*$; and the electron mean free path is much shorter than the
parallel length of the device, $\lambda_{ei} \ll L_{||}$.  Therefore
for low frequency, long parallel wavelength modes a collisional fluid
model might be reasonable choice for modeling LAPD plasmas.  Kinetic
effects can, however, be very important in LAPD, in particular for
Alfv\'{e}n waves, where the electron thermal speed is comparable to
the phase speed of the wave, $v_\phi \sim v_{\rm
  th,e}$~\cite{Vincena2001}. Because of the large parallel size of
LAPD (and plasma beta of order the mass ratio, $\beta \sim m_e/M$),
drift waves couple to Alfv\'{e}n waves~\cite{maggs2003} and kinetic
effects may be important~\cite{penano1997}.  It can be argued that
even in this case strong collisions may disrupt kinetic processes such
as Landau damping and fluid description may still provide a good
description of the plasma~\cite{ono1975}. Nonlinear BOUT simulations
using fluid equations can help to identify the limits of the validity
of the collisional fluid model in LAPD.

For the simulations described here the following set of
equations are used:
\beqar
\label{eqBN}
\pdt N & = & - \vE\cdot\grad N - \gradpar(\vpe N)\\
\label{eqBvpar}
\pdt\vpe & = & - \vE\cdot\grad \vpe -
\mu\frac{T_{e}}{N}\gradpar N + \mu\gradpar\phi - \nue\vpe\\
\label{eqVort}
\pdt\varpi & = &
-\vE\cdot\grad\varpi 
- \gradpar(N\vpe) 
+ \bvec\times\grad N\cdot\grad\vE^2/2
-\nuin\varpi
+\mu_i \gradperp^2\varpi
\eeqar
Here $N$ is the plasma density, $\vpe$ is the electron fluid parallel
velocity, and $\varpi$ is the potential vorticity introduced as 
\beq\label{eqDefVort}
\varpi\defeq\gradperp\cdot\left(N\gradperp\phi\right) 
\eeq
% Equations~(\ref{eqBN}-\ref{eqDefVort}) are described in detail
elsewhere~\cite{Popovich2010a} except the viscosity term that is added
for nonlinear calculations and is discussed in
Appendix~\ref{secAppendixVisc}.  All the quantities here are
normalized using the Bohm convention.  The model used in BOUT is
similar to that employed in other efforts to simulate linear devices,
in particular on LMD~\cite{kasuya2007}, CSDX~\cite{Holland2007}, and
the recent work by Rogers and Ricci in simulating turbulence in
LAPD~\cite{Rogers2010}.

Density, temperature and magnetic field are normalized to reference values
$n_x$, $T_{ex}$ (the maximum of the corresponding equilibrium
profiles), and $B_0$, the axial magnetic field. Frequencies and time
derivatives are normalized to $\wcix = eB_0/m_ic$: $\hat{\pdt} =
\pdt/\wcix$, $\hat{\omega} = \omega/\wcix$; velocities are normalized
to the ion sound speed $C_{sx}=\sqrt{T_{ex}/m_i}$; lengths -- to the
ion sound gyroradius $\rho_{sx} = C_{sx}/\wcix$; electrostatic
potential to the reference electron temperature: $\hat{\phi} =
e\phi/T_{ex}$.  In Eqs.~(\ref{eqBN}-\ref{eqDefVort}) and further, the
$''~\hat{~}~''$ symbol for dimensionless quantities is dropped
for brevity of notation.

While the variables $N$, $\vpe$ and $\varpi$ are advanced in time,
equation (\ref{eqDefVort}) is solved to reconstruct the perturbed
potential $\phi$ from $\varpi$. In the code version used in this work,
Eq.~(\ref{eqDefVort}) is linearized to increase for computational efficiency,
since this equation has to be solved for each evaluation of the
right-hand side of Eqs.~(\ref{eqBN}-\ref{eqVort}).  The vorticity
evolution equation (\ref{eqVort}) replaces the current continuity
equation. Derivation and discussion of this form of equation is
presented elsewhere~\cite{Popovich2010a}. Derivation of the viscosity term is
discussed in Appendix~\ref{secAppendixVisc}.

Equations~(\ref{eqBN}-\ref{eqDefVort}) do not include perturbations of
the magnetic field. While electromagnetic effects are essential to
capture the physics of Alfv\'{e}n and drift-Alfv\'{e}n waves,
preliminary linear simulations indicate that the effect of magnetic
perturbations on frequencies and growth rates for low-frequency drift
wave is small for the LAPD parameters considered here. A more
detailed nonlinear study of the electromagnetic physics is a subject
of future work and is outside the scope of this study.

Time evolution equations (\ref{eqBN}-\ref{eqDefVort}) are implemented
in the numerical code BOUT~\cite{Xu1998,Umansky2009}. Originally
developed for simulations of the tokamak edge plasma, the code has
been adapted to the cylindrical geometry of LAPD. For the present
study special care is taken to avoid spurious numerical solutions due
to discretization in the coordinate parallel to the magnetic field
(see Appendix~\ref{secAppendixGrid}). Prior to the turbulence
calculations presented here the code has been successfully verified for a
range of linear instabilities potentially existing in the LAPD plasma,
including the resistive drift, Kelvin-Helmholtz and rotational
interchange instabilities~\cite{Popovich2010a}.

%%%%%%%%%%%%%%%%%%%%%%%%%%%%%%%%%%%%%%%%%%%%%%%%%%%%%%%%%%%%%%%%%%%%%%%%%%%

\section{Turbulent transport and average density profile}\label{secAvgProfile}

\subsection{Average and local fluctuating fields}

\begin{figure}[htbp]
\begin{center}
\includegraphics[width=0.7\textwidth]{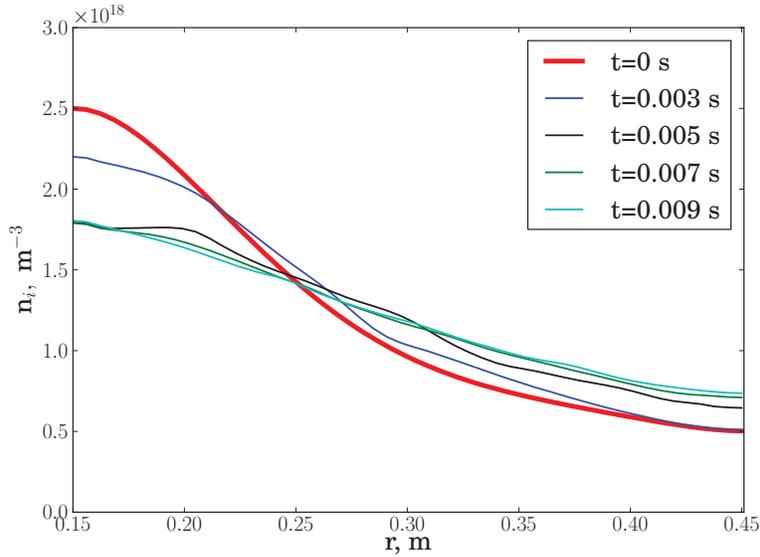}
\end{center}
\caption{\label{fig_ni_evolution}(Color Online) Relaxation of the density profile
in a simulation without sources.}
\end{figure}

In turbulence where the eddy size is much smaller than the macroscopic
system size, the separation of spatial scales usually leads to separation of
time-scales for the evolution of local and spatially averaged
fields. In spite of this separation of scales, the average and
fluctuating quantities are certainly coupled since 
gradients provide the source of free energy driving turbulence; on the
other hand, turbulent transport, along with sources, leads to
evolution of the macroscopic profiles. If no sources are present in
the simulation, the profiles relax to smaller gradient as is shown in 
Fig.~\ref{fig_ni_evolution}.  For comparison of the simulated and
measured turbulence characteristics, this profile evolution is usually
undesired, since this comparison requires collecting a large
statistical sample of data for stationary, experimentally relevant
profiles.

For the purposes of considering algorithms for average profile control in
BOUT, it is convenient to represent the fluctuating variables at any spatial
location as a sum of the time-average and perturbation,
\beq
f(\xvec,t)=\bar{f}(\xvec)+\tilde{f}(\xvec,t),
\eeq
For cylindrically symmetric configuration it is also useful to
separate $f$ into azimuthally symmetric and asymmetric components,
\beq
f(\xvec,t)=\savg{f(\xvec,t)}+\{f(\xvec,t)\},
\eeq

where $\savg{f} = (1/2\pi)\int f d\theta$ and $\{f\}=f-\savg{f}$ is the
residual.   Based on the ergodic hypothesis, it is assumed that the time-average
$\bar{f}$ is equal to the azimuthal-average $\savg{f}$, and
statistical moments of $\tilde{f}$ and $\{f\}$ are equal. This separation
of variables into an axisymmetric and non-axisymmetric part does
not preclude a full nonlinear solution in BOUT, but it allows easier control
over the average profiles of density, temperature and other quantities.

\subsection{Profile maintenance: suppressing the azimuthal average}\label{secAvgProfile_zavg}

Following self-consistent time evolution of turbulence and macroscopic
transport may be difficult because the time-scale separation can make
such calculations too large to be practical. Additionally,
including first-principles-based source terms, e.g., for density and
temperature, can be complicated, involving models for the plasma
source, atomic physics, radiation transport, etc. 

\begin{figure}[htbp]
\begin{center}
\includegraphics[width=0.8\textwidth]{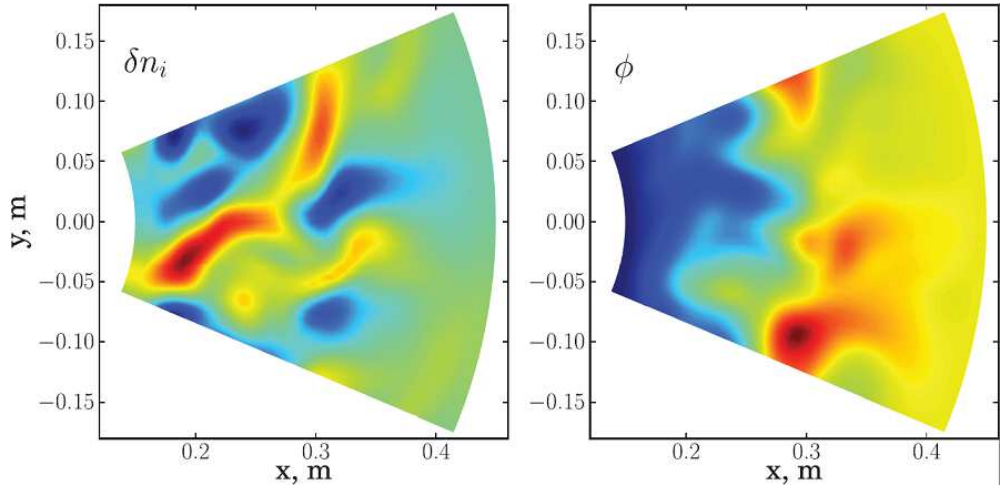}
\end{center}
\caption{\label{fig_turb2d}(Color Online) Spatial structure of the perturbed density (left) and turbulence-generated potential (right) at $t=5.2{\rm~ms}$.}
\end{figure}

\begin{figure}[htbp]
\begin{center}
\includegraphics[width=0.7\textwidth]{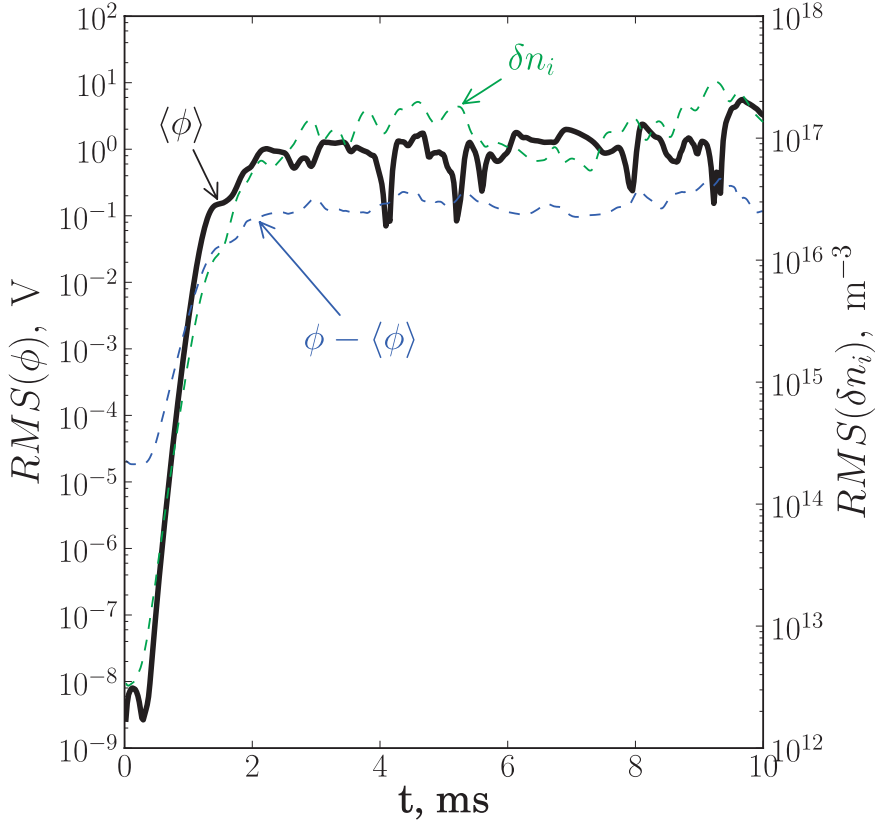}
\end{center}
\caption{\label{fig_rms}(Color Online) Time evolution of the density and potential fluctuation RMS in a typical simulation. The potential is split into the axisymmetric ($\left<\phi\right>$, zonal flow component) and the non-axisymmentric ($\phi-\left<\phi\right>$) components.}
\end{figure}

Without attempting a self-consistent time evolution of turbulence and
macroscopic transport one can consider intermediate time-scales $\tau
\ll t \ll T$, where the macroscopic profiles can be taken as
``frozen'', based on known measured experimental average profiles. In
this case the time-evolution of only the non-axisymmetric part is
considered, and a simple technique of maintaining the desired average
profile is filtering out the axisymmetric part of evolving fields.
This is illustrated in Fig.~(\ref{fig_turb2d}-\ref{fig_rms}) showing
the general appearance of $\phi$, $\delta n_i$, and the evolution of
the density and potential fluctuation RMS in a simulation with
``frozen'' density profile. In Fig.~(\ref{fig_rms}), the potential is
split into the turbulence-generated axisymmetric component
$\left<\phi\right>$, and the non-axisymmetric residual
$\{\phi\}=\phi-\left<\phi\right>$. One can observe the development of a
zonal flow component $\savg{\phi(r)}$ corresponding to sheared
azimuthal flow.

The easiest way to control the average profile is to suppress the
evolution of the axisymmetric component by subtracting the azimuthal
average of the right-hand side of Eqs.~(\ref{eqBN}-\ref{eqVort});
for example, for the density:
\beq
\pdt N = RHS - \thavg{RHS}
\eeq
This is effectively introducing a time-dependent source/sink function
necessary to maintain exactly the target density profile.

However, suppressing the axisymmetric part of fluctuations may interfere too
much with the solution, and one can consider doing this not in the
full domain but only on the boundary. This would constrain the
boundary values, and may be enough to maintain the average profile
close to the desired. 

\subsection{Profile maintenance through adding sources}\label{secAvgProfile_Source}

A more physical method to control the profile is to use a source/sink
term $S(r)$ designed in such a way that the average profile is
maintained close to the experimentally measured profile.  Rather than
developing this source from first principles, an {\itshape ad hoc} source/sink is
chosen in order to achieve the desired steady state profile in the
simulation.  As a first step, a simulation is performed using the
method of subtracting out the azimuthal average to maintain the
$\savg{N_i}$ profile close to the ``target'' density profile
${N_{i0}(r)}$, which is based on a representative experimental probe
measurements in LAPD. Once a steady-state turbulence solution is
obtained, the radial particle flux is calculated from fluctuating
density and potential,
\beq
\Gamma = \savg{N_i V_{Er}}
\eeq

Next, the effective volumetric density source $S(r)$ is calculated as
\beq
S =\grad \cdot \Gamma
\eeq
that can now be added to the density evolution equation, Eq.~(\ref{eqBN}):
\beq
\pdt N = RHS + S(r)
\eeq

\begin{figure}[htbp]
\begin{center}
\includegraphics[width=0.7\textwidth]{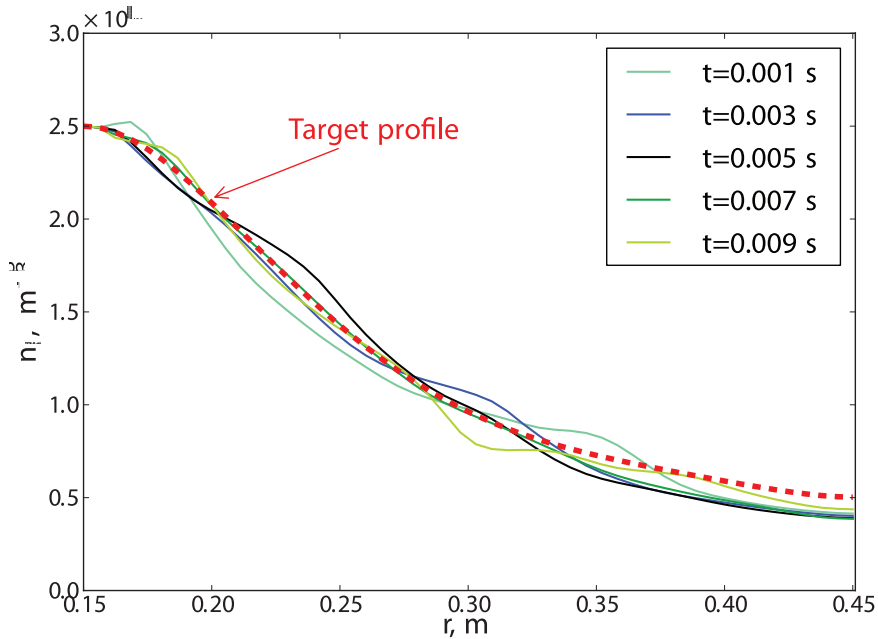}
\end{center}
\caption{\label{fig_ni_wsource}(Color Online) Instantaneous density profile
$\savg{N_i}$ in simulation with density source $S(r)$.}
\end{figure}

\begin{figure}[htbp]
\begin{center}
\includegraphics[width=0.7\textwidth]{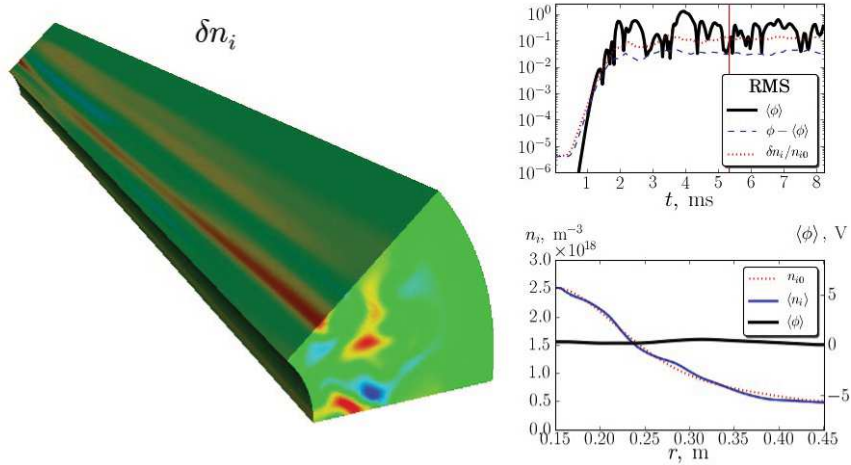}
\end{center}
\caption{\label{fig_ni_animation}(Color Online) Evolution of density and potential
in BOUT simulation with density source.  Left: density fluctuations;
right: RMS of the density perturbations and the axisymmetric and
non-axisymmetric components of self-generated potential (top); average
radial profiles of density and potential (bottom) (enhanced online).[URL: http://dx.doi.org/10.1063/1.3527987.1]}
\end{figure}

A subsequent simulation is run with this new source term and without
suppressing the azimuthal average, allowing turbulent transport to
compete with the source term.  If needed, another iteration can be
made by adjusting the source term to account for the mismatch between
the BOUT predicted profile and the target. A more comprehensive
approach to self-consistent time-evolution of turbulence and average
profiles can be based on adding an adaptive source~\cite{Dudson2009};
however it is beyond the scope of this paper. In the present study
extra iterations were not necessary; a single step was enough to
produce stationary turbulence with average density profile close to
the target profile, as shown in Fig.~\ref{fig_ni_wsource}. The
evolution of the density and potential in a typical BOUT simulation
with density source and fixed values of the density at the radial
boundary is presented in Fig.~\ref{fig_ni_animation} (animation online).

%%%%%%%%%%%%%%%%%%%%%%%%%%%%%%%%%%%%%%%%%%%%%%%%%%%%%%%%%%%%%%%%%%%%%%%

\section{Comparison with LAPD data}\label{secCompare}

Before any attempt is made to construct a full simulation of LAPD that
self-consistently incorporates transport, first principles
sources/sinks and profile evolution, it is necessary to ensure that
the basic characteristics of the turbulence are correctly captured by
the physical model being used.  Initial simulations have been done
using periodic axial boundary conditions and using the experimentally
measured density profile.  A constant temperature profile (5 eV) is
used and the potential is allowed to evolve self-consistently
(potential and flow profiles are not matched to experimentally
measured profiles).  The main experimental dataset used in this work
is taken from Carter~\etal.~\cite{carter2009}, and includes
measurements of plasma and turbulence profiles and two-dimensional
correlation functions.  Density fluctuations in BOUT are compared
against ion saturation current $\Isat$ fluctuations measured in the
experiment (making the currently unverified assumption that
temperature fluctuations are negligible).

To be able to compare turbulence characteristics to the measured data
for the experimentally relevant plasma parameters, the average
profiles have to be maintained close to the experimental values during
the simulation. To do so, the two methods
described in section~\ref{secAvgProfile} are applied, either subtracting the
azimuthal average of the right-hand side of the density equation, or
adding a time-independent source function $S(r)$ to the right-hand
side of Eq.~(\ref{eqBN}).

Using these two methods, a steady-state turbulence for LAPD parameters
is simulated, solving Eqs.~(\ref{eqBN}-\ref{eqDefVort}) with the
average density close to the experimental profile, for a range of
ion-neutral collisionalities ($\nuin/\Omega_{ci}=2\times10^{-4},
1\times10^{-3}, 2\times10^{-3}$). The estimated value for LAPD, based
on neutral density $n_n\sim5\times10^{11}cm^{-3}$, is
$\nuin/\Omega_{ci}=2\times10^{-3}$.  The simulations are performed in
the radial interval $0.15\le r\le 0.45{\rm~m}$, in an azimuthal
segment of a cylinder of $\pi/4$ angle, assuming periodicity in the
azimuthal angle and parallel direction.  The grid size is
$50\times32\times32$ points for radial, azimuthal and parallel
coordinates. In order to improve the statistics, a series of
uncorrelated runs is made with slightly different initial
perturbations. In the experiment, a slow evolution of the mean density
is observed on a $\sim 1-2\rm{~ms}$ time scale. BOUT simulations with
fixed background profile can not capture the slow variation of the
average density because the azumuthally symmetric component of density
perturbation is continuously removed in BOUT simulation to maintain a
constant profile. For the purposes of direct
comparisons between BOUT results and the measured data, this slow
evolution is removed from the experimental data by applying a temporal
smoothing of the signal, which cuts off all frequencies below
800~Hz. For consistency, the same smoothing is applied to BOUT data.

\subsection{Fluctuations: temporal and spatial characteristics, PDF}\label{Skew}

\begin{figure}[htbp]
\begin{center}
\includegraphics[width=0.7\textwidth]{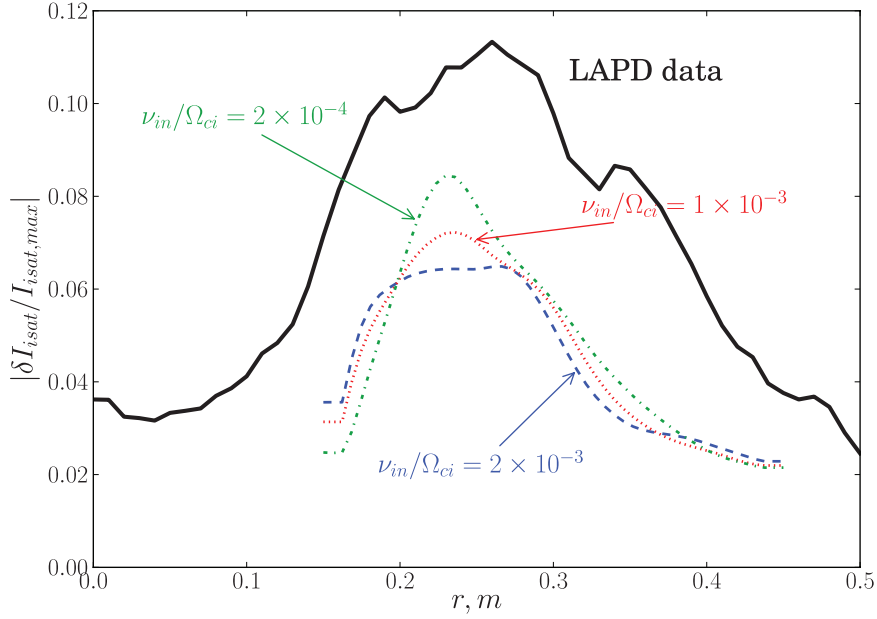}
\end{center}
\caption{\label{fig_AMP}(Color Online) Radial distribution of the average ion saturation
current fluctuations normalized to the maximum equilibrium profile value.}
\end{figure}

The analysis of BOUT results shows that the $\Isat$ fluctuation
amplitudes in LAPD data and the simulations have similar radial
location near the cathode edge ($r\sim 28~\cm$), where the background
density gradient is largest, as shown in Fig.~\ref{fig_AMP}. The
absolute values of the fluctuations are of the same order of
magnitude, with simulated amplitudes smaller by a factor $\lesssim$2
than the experimental data.    

\medskip

\begin{figure}[htbp]
\begin{center}
\includegraphics[width=0.7\textwidth]{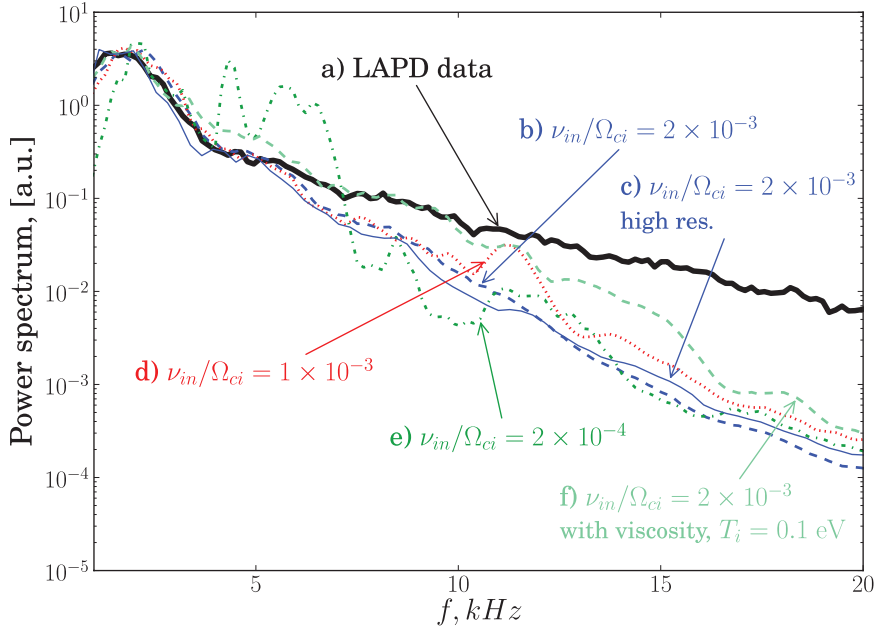}
\end{center}
\caption{\label{fig_FFT}(Color Online) Frequency power spectrum of the density
fluctuations: LAPD measurements (a) and BOUT simulations (b-f) for
b)~$\nuin/\wci=2\times10^{-3}$, c)~$\nuin/\wci=2\times10^{-3}$ with
$N_z=128$ azimuthal grid size, d)~$\nuin/\wci=1\times10^{-3}$,
e)~$\nuin/\wci=2\times10^{-4}$, f)~$\nuin/\wci=2\times10^{-3}$ with
ion viscosity at $T_i=0.1{\rm~eV}$. Experimental density profile,
$T_e=5eV$.}
\end{figure}

The comparison of the frequency power spectrum of the density
fluctuations $\delta n/n$ to the LAPD measured spectrum is presented
in Fig.~\ref{fig_FFT}. The spectra are integrated over the volume
$0.22 \le r\le 0.28~{\rm m}$, using a sliding Hanning window for
averaging between the different simulation runs. Note that the total
power in each spectrum is normalized to obtain the best fit to the
experimental data.  The power spectral shape from the BOUT simulation
(for a range of $\nuin$ values) is in a relatively good agreement with
the measured spectrum (Fig.~\ref{fig_FFT}, b,d,e). At higher
frequencies, $\gtrsim 10{\rm~kHz}$, the simulated spectra fall off
faster than the measured spectrum. More studies are required to
analyse the effect of additional features of the physical model
(temperature profile and perturbations, sheath boundary conditions,
etc.) on the power spectrum, as well as the numerical resolution. High
frequencies corresponding to smaller spatial structures are
potentially more affected by finite resolution effects.
As a check of numerical convergence in terms of box size, the spectrum for
$\nuin/\Omega_{ci}=\enum{2}{-3}$ is calculated with four times the
size of the azimuthal extent of the grid and the number of azimuthal
grid points ($N_z=128$). The spectrum shape is similar to the original
calculation, with a smaller grid size (Fig.~\ref{fig_FFT}, b and c).
These simulations were performed without the explicit ion viscosity
term in the vorticity equation~(\ref{eqVort}), therefore the only
viscosity is due to the numerical discretization. Inclusion of the
ion-ion viscosity term as discussed in the
Appendix~\ref{secAppendixVisc} corresponding to the ion temperature
$T_i=0.1\eV$ does not significantly change the spectrum shape.
The effects of viscosity in BOUT simulations is
a subject of ongoing work.  It is interesting to note that both the
experimental and the simulation spectra exhibit an exponential power
spectrum at higher frequencies (straight line on the log-lin plot),
which is consistent with the presence of coherent structures~\cite{pace2008}.

\begin{figure}[htbp]
\begin{center}
\includegraphics[width=0.7\textwidth]{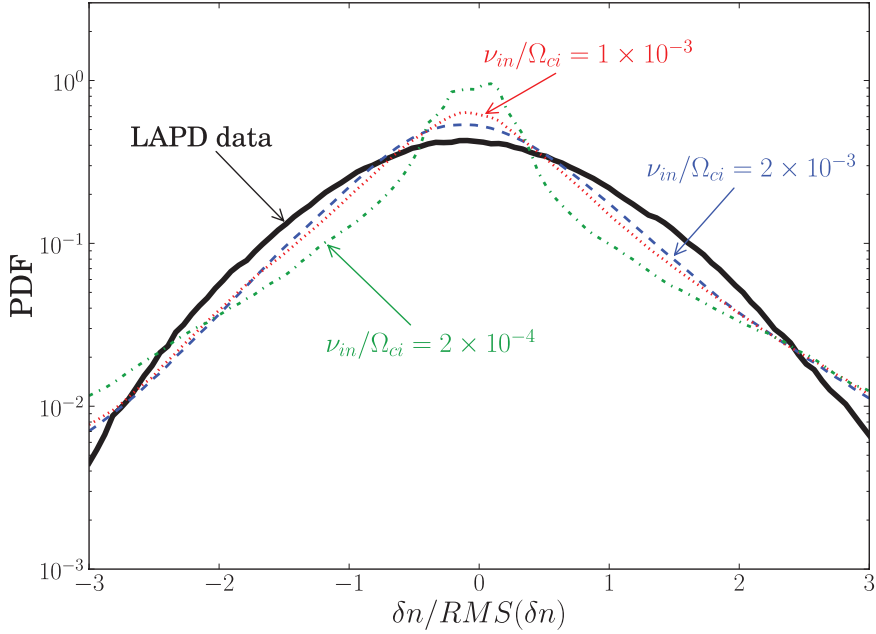}
\end{center}
\caption{\label{fig_PDF}(Color Online) Probability distribution function of
$\delta n/RMS(\delta n)$ fluctuation amplitude in BOUT simulations and LAPD
data. PDF is volume averaged in the interval $0.22\le r\le
0.28{\rm~m}$.}
\end{figure}

\begin{figure}[htbp]
\begin{center}
\includegraphics[width=0.7\textwidth]{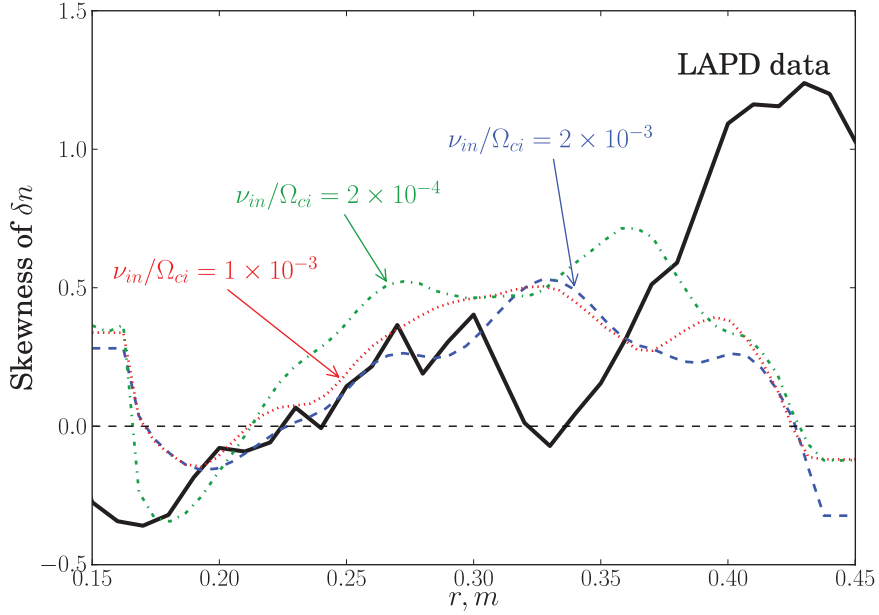}
\end{center}
\caption{\label{fig_SKEW}(Color Online) Skewness of $\delta n$ distribution in
BOUT simulations and LAPD measurements.}
\end{figure}

Another important characteristic of the turbulence is the probability
distribution function (PDF) of fluctuation amplitudes. The PDF of
$\delta n/RMS(\delta n)$ fluctuations from LAPD probe data is integrated over a
volume of plasma $0.22\le r\le0.28~{\rm m}$ and compared with the PDF
from BOUT simulations in the same volume (Fig.~\ref{fig_PDF}). There
are no normalizations or fit factors involved in this comparison. The
experimental and the simulated PDFs are similar, with the average
relative fluctuation $\savg{|\delta n/n|}$ of 0.16 for the measured
data and 0.09, 0.09, 0.08 for BOUT simulations with
$\nuin/\Omega_{ci}=\enum{2}{-4}, \enum{1}{-3}, \enum{2}{-3}$. The PDF
for $\nuin/\Omega_{ci}=\enum{2}{-3}$ case is the closest to the
experimental data, which is consistent with the estimate of the
neutral density in LAPD. Note that the distribution is mostly
symmetric here because it is integrated over a radial inteval 
where the skewness is relatively low (Fig.~\ref{fig_SKEW}).

Intermittent turbulence is observed in the edge plasmas of many
experimental devices. This intermittency is usually attributed to
generation and transport of coherent filaments of plasma, ``blobs'' or
``holes''~\cite{Carter2006}. One signature of the presence of these
structures is the non-zero skewness of the density fluctuation
PDF. Typically, positive skewness associated with convective transport
of ``blobs'' is observed in LAPD measurements in the region outside of
the cathode edge ($r \gtrsim 28~\cm$). Smaller negative values,
associated with the ``holes'' are usually observed inside the cathode
radius. The radial profile of the skewness of $\delta n$ is shown in
Fig.~\ref{fig_SKEW}. Except for the edge of the simulation domain
which is affected by the imposed boundary conditions, the trend of the
skewness profile, as well as the absolute values, is similar in BOUT
simulations and in the LAPD data.

\begin{figure}[htbp]
\begin{center}
\includegraphics[width=0.8\textwidth]{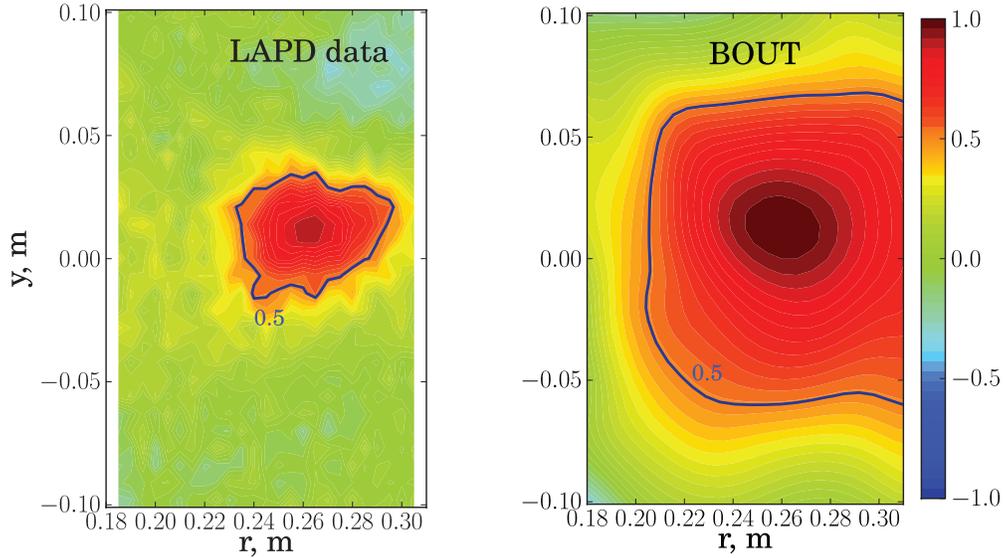}
\end{center}
\caption{\label{fig_dnCorr}(Color Online) Correlation function for $\Isat$
fluctuations measured using a moving probe.}
\end{figure}

Two-dimensional turbulent correlation functions are measured in LAPD
using two probes: a fixed reference probe and a second probe that is
moved shot-to-shot to many ($\sim$ 1000) spatial locations in a 2D
plane perpendicular to the magnetic field. The
reference probe remains at a fixed position that is close enough to
the moving probe in the axial direction so that the parallel variation
of the turbulent structures can be neglected. This allows to obtain
the 2D spatial correlation function in the azimuthal plane. 
A similar ``synthetic'' diagnostic to post-process BOUT
simulation results is constructed by calculating the correlations between a reference
location and all other points in each azimuthal plane. The correlation
length in BOUT simulation is of the same order, but larger than the
measured value (Fig.~\ref{fig_dnCorr}).  

\subsection{Fluxes and sources vs. inferred source/sink in LAPD}

\begin{figure}[htbp]
\begin{center}
\includegraphics[width=0.7\textwidth]{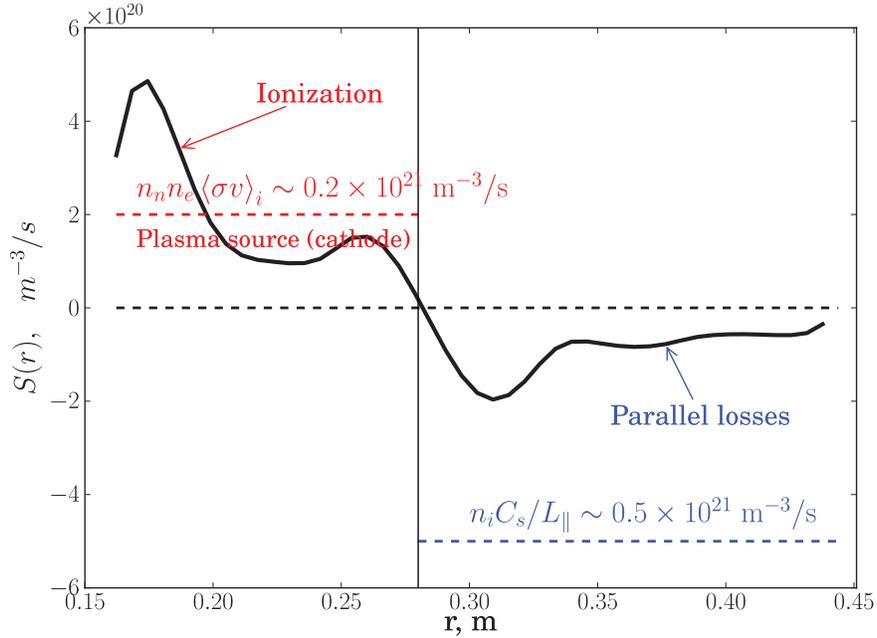}
\end{center}
\caption{\label{fig_ni_source}(Color Online) Inferred particle source required
to maintain the measured density profile.}
\end{figure}

An inferred particle source is required to maintain the density
profile close to the experimental values, as described in
section~\ref{secAvgProfile_Source}. In BOUT simulations, the
calculated source function $S(r)$ that produces steady-state
turbulence with the desired density profile appears to be positive
inside $r_0 \sim 28~\cm$, and negative outside
(Fig.~(\ref{fig_ni_source})). Remarkably, the qualitative form and
magnitude of the $S(r)$ profile is consistent with the assumption that
within $\sim r_0$ there is an ionization source, and outside of $\sim
r_0$ there is a sink due to parallel streaming to the end walls. In
LAPD, the field lines inside $r\sim 28$~cm connect to the cathode that
produces the primary ionizing electron beam. The magnitude of the
inferred source is close to the estimated ionization source and
parallel losses sink for LAPD plasma, $S_{source}\sim n_n n_e <\sigma
v>_i\sim\enum{2}{20}~\mn/s$ and $S_{sink}\sim n_i
C_s/L_{\parallel}\sim\enum{5}{20}~\mn/s$.

\section{Discussion}\label{secDiscussion}

\subsection{Plasma transport}

\begin{figure}[htbp]
\begin{center}
\includegraphics[width=0.8\textwidth]{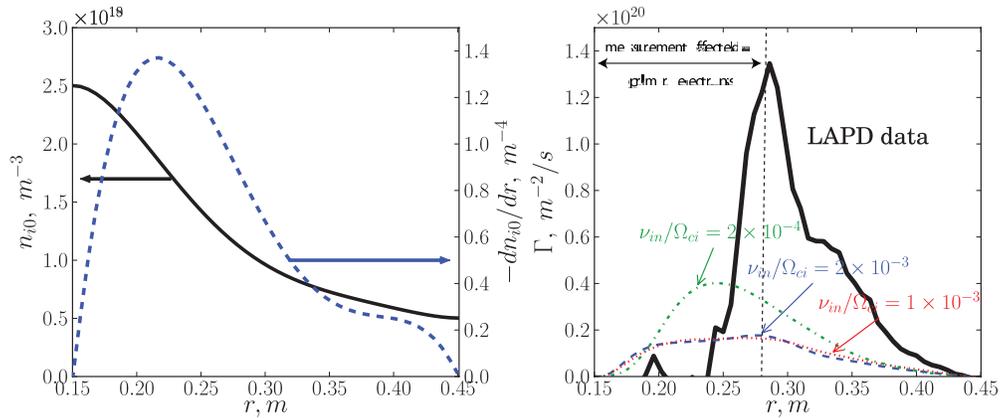}
\end{center}
\caption{\label{fig_dcoef}(Color Online) Left: Radial profiles of average $n_i$ and $dn_i/dr$. Right: radial particle flux $\Gamma$ from the measured LAPD data and BOUT simulations.}
\end{figure}

One can calculate the effective diffusion coefficient by dividing the
radial flux by the gradient of the equilibrium density,
\beq
D_{eff} = - \frac{\Gamma}{\grad N_{i0}}
\eeq
The turbulence-driven radial particle flux in BOUT simulations peaks
near the maximum density profile gradient and is close to diffusive
model value, with the effective diffusion coefficient $D_{eff}\sim
3{\rm~m^2/s}$ on the order of Bohm value, $D_{Bohm}\sim 8{\rm~m^2/s}$.
The profiles of the average radial flux measured in LAPD and
calculated in BOUT are shown in Fig.~(\ref{fig_dcoef}).  The simulated
flux is significantly lower than the measured flux in LAPD, but this
difference is consistent with the difference in fluctuation amplitude
shown in Fig.~\ref{fig_AMP}. It should also be noted that the measured
flux profile is valid only for $r > 28$cm as the measurement of
azimuthal electric field fluctuations in LAPD is affected by fast
electrons generated by the plasma source~\cite{carter2009}.

While comparisons are made here to a diffusive model and particle
transport in LAPD has been shown to be well modeled by Bohm
diffusion~\cite{maggs2007}, the simulated plasma turbulence has
features that are usually associated with intermittency, as discussed
in section~\ref{Skew}: non-Gaussian fluctuations, as seen by non-zero
standard statistical moments.  The role of coherent structures and
convective transport in BOUT simulations is the subject of ongoing
investigation.

%	Numerical diagnostics and Er equation
\subsection{Numerical diagnostics and $\savg{E_r}$ equation}

An important part of the simulation effort is the framework for
various diagnostics of the numerical solution. An obvious test of the
solution is the check of the conservation laws (particle
number, momentum, etc). However, depending on the choice of the radial boundary
conditions and the form of the sources, the total number of particles,
for example, is not necessarily conserved during the simulation. 
A more appropriate diagnostics in this case is the local check of the
balance of the terms in each of the time evolution equations
(\ref{eqBN}-\ref{eqVort}), at each point in space and time. As well as
ensuring that the solution is correct, calculation of this balance
also provides a useful insight into the relative importance of the
physical terms. 

This test can be done directly, using the explicit form of the
equations (\ref{eqBN}-\ref{eqVort}) as they appear in the code, or
indirectly, by calculating the conservation of physical quantities not
directly solved for by BOUT. For the direct tests, the balance of the
terms in BOUT equations is satisfied to the machine accuracy when the
appropriate finite difference schemes are used in the diagnostic
module.

Tests involving equations not directly solved by BOUT can be
more subtle. An example is the equation for the azimuthal
momentum. BOUT simulations in LAPD geometry show generation of
self-consistent axisymmetric component of the potential. The dynamics
of this zonal flow component can be derived from the vorticity
equation as shown in Appendix~\ref{secAppendixEr}:
\beq\label{eqmv}
\pdt\savg{N V_\theta} 
=
 \savg{\varpi V_r} 
+ \frac{1}{r}\savg{N\pdth\frac{\gradperp\phi^2}{2}}
- \nuin\savg{N V_\theta} 
- \mu\pdr\savg{\varpi} 
\eeq
This expression is similar to the equation for the zonal flow
component of the radial electric field that can be derived directly
from the azimuthal projection of the ion momentum equation
\cite{Tynan2006}:
\beq\label{eqEr}
\pdt\savg{V_\theta} 
=
-\frac{1}{r^2}\pdr\left(r^2\savg{V_rV_\theta}\right)
- \nuin\savg{V_\theta} 
+ \mu\left(\pdr^2\savg{V_\theta}+\frac{1}{r}\pdr\savg{V_\theta}-\frac{1}{r^2}\savg{V_\theta} \right) 
\eeq
The first term in the right-hand side
of Eq.~(\ref{eqmv}) and Eq.~(\ref{eqEr}) is the turbulent Reynolds
stress. If a stationary turbulent state exists, the time average of
the Reynolds stress, which is the driving term for the zonal flows, is
balanced by the ion-neutral collisions and the ion viscosity terms.
This equation is not derived directly from the vorticity equation,
therefore BOUT solution satisfies it only to the extent that the
underlying assumptions in the derivation of the vorticity equation
(\ref{eqVort}) are fulfilled. 

Applying this diagnostic to BOUT output, it is found that the balance is
well satisfied (within $\sim 5\%$) for Eq.~(\ref{eqmv}) (with a small
correction due to the linearization used in the inversion of
Eq.~(\ref{eqDefVort})) and Eq.~(\ref{eqEr}) with a particular choice
of the finite difference scheme in the advection operators -- the
fourth order central scheme. However, for the first order upwind
scheme used in most of nonlinear simulations presented here, the balance is not
sufficiently well satisfied, which indicates that a higher numerical
resolution is required to reach convergence for this diagnostic measure.

At present, a quantitative match between the
average $\savg{E_r}$ in the code and in the experimental
data has not been obtained. However, at present the model does not include all physics
(e.g., sheath, biasing, temperature perturbations) that is certainly
important for setting the average radial electric field. Improving the model by adding
to it the missing physics to address matching of $\savg{E_r}$ is the
subject of ongoing research.

%e.	Future work
\subsection{Future work}

A detailed verification study of linear instabilities in LAPD using
BOUT~\cite{Popovich2010a} combined with a good qualitative and even
partially quantivative agreement between nonlinear BOUT simulations
and LAPD measuments presented here provides confidence in the
relevance of these simulations to LAPD. It is, however, not a fully
consistent first-principle model at present, and some potentially
important physics is yet to be included. Most importantly, the
experimentally measured temperature and flow profiles need to be
matched and an evolution equation for temperature fluctuations,
already implemented in BOUT, could be employed.  The addition of a
temperature gradient to the simulation would likely increase
instability drive and result in a larger saturated turbulent amplitude
and particle flux, more consistent with observation.  
Another significant improvement of the model that remains to be implemented is
the sheath boundary condition at the end plates in the parallel
direction. The axial boundary conditions are expected to be important
in the formation of the radial electric field profile in LAPD, and a
more physical description might help to understand and model the
dynamics of the azimuthal flows and experimentally relevant potential
profiles. The role of the ion viscosity on these flows is also a
subject of ongoing work and has to be investigated in more
detail. Additionally, including sheath boundary conditions along with
temperature gradients, electron temperature
fluctuations and an equation for temperature evolution can give rise
to modifications to drift instabilities and introduce new modes, such
as entropy waves~\cite{Tsai1970,ricci2006} and conducting wall
modes~\cite{Berk1991}. 

Although the linear calculations indicate that electromagnetic effects
do not significantly affect drift wave instability in LAPD, magnetic
field perturbations are required for Alfv\'en wave studies. Alfv\'en
waves represent an important part of LAPD research, and nonlinear
simulations of Alfv\'en waves using BOUT can contribute to the
understanding of electromagnetic turbulence in LAPD.

%%%%%%%%%%%%%%%%%%%%%%%%%%%%%%%%%%%%%%%%%%%%%%%%%%%%%%%%%%%%%%%%%%%%%%%%%%%

\section{Conclusions}\label{secConclusions}

A numerical 3-D model of plasma turbulence is applied to LAPD. The
physics model includes equations for plasma density, electron parallel
momentum, and current continuity, for partially ionized plasma. The
model is implemented in the numerical code BOUT that is adapted for
cylindrical geometry. This model has previously been successfully
verified for a range of linear instabilities in LAPD~\cite{Popovich2010a}.

Two different methods for average profile control are applied in the
simulations presented in this study. One approach consists of
suppressing the azimuthal average of the density fluctuations by
directly subtracting it from the time evolution equation. The second
method uses a source/sink radial function that is constructed to balance
steady-state radial transport flows. Both methods successfully maintain the
average density profile close to the experimental value, required for
comparisons with the measured turbulence characteristics.

Nonlinear BOUT simulations demonstrate a self-consistent evolution of
turbulence and self-generated electric field and zonal flows, that
saturates and results in a steady turbulent state. The simulated
fluctuation amplitudes in the steady state are within a factor of 2 of
the measured data, with a similar radial location near the cathode
edge. The probability distribution function of the fluctuation
amplitudes is comparable to the experimental distribution.
Statistical properties of edge turbulence, such as the skewness, are
often used as indication of the turbulence
intermittency~\cite{zweben2007}.  In tokamak edge, the skewness is
positive in the SOL (i.e. dominated by large amplitude events), but is
sometimes negative inside the separatrix or limiter radius (i.e. with
density ``holes''), e.g. seen in DIII-D~\cite{Boedo2003} and
NSTX~\cite{Boedo2001}. This feature is similar to LAPD data, and
reproduced in BOUT simulations. Despite the intermittent character of
turbulence, as indicated by non-Gaussian PDF, the turbulent particle
flux magnitude is consistent with diffusive model with diffusion
coefficient on the order of $D_{Bohm}$. The inferred particle source/sink
function required to maintain the simulated steady-state density
profile close to experimental value is consistent with the estimates
of ionization sources and parallel losses in LAPD discharge.

The spatial and temporal structures of
the fluctuations are consistent with the LAPD measurements,
however the correlation length in BOUT simulations is
larger than in the experiment.

Although some elements of the physical model still remain to be
implemented (sheath parallel boundary conditions, azimuthal flow
matching, electron temperature fluctuations, etc.), the agreement
between certain features of the experimental data and simulations
based on this relatively simple model lends confidence in the
applicability of these simulations to the LAPD plasmas.

%%%%%%%%%%%%%%%%%%%%%%%%%%%%%%%%%%%%%%%%%%%%%%%%%%%%%%%%%%%%%%%%%%%%%%%%%%%

\begin{acknowledgments}

This work was supported by DOE Fusion Science Center Cooperative
Agreement DE-FC02-04ER54785, NSF Grant PHY-0903913, and by LLNL under
DOE Contract DE-AC52-07NA27344.  BF acknowledges support through
appointment to the Fusion Energy Sciences Fellowship Program
administered by Oak Ridge Institute for Science and Education under a
contract between the U.S. Department of Energy and the Oak Ridge
Associated Universities.

\end{acknowledgments}

\appendix
%\appendixpage

%%%%%%%%%%%%%%%%%%%%%%%%%%%%%%%%%%%%%%%%%%%%%%%%%%%%%%%%%%%%%%%%%%%%%%%%%%%

\section{Azimuthal momentum equation}\label{secAppendixEr}

The equation describing the evolution of the surface-averaged
azimuthal momentum can be derived from the vorticity equation:
\beq
\label{eqVortA}
\pdt\varpi =
-\vE\cdot\grad\varpi 
- \gradpar(N\vpar) 
+ \bvec\times\grad N\cdot\grad\vE^2/2
-\nuin\varpi
+ \mu_i \gradperp^2 \varpi
\eeq
where the potential vorticity is
\beq\label{eqDefVortA}
\varpi = \gradperp\cdot\left(N\gradperp\phi\right) 
\eeq

Define surface and volume averaging by 
\beq
\savg{f(r,\theta,z)} = \frac{1}{2 \pi L}  \int_0^{2\pi}\int_0^{L} {f dz d\theta}
\eeq
\beq
\vavg{f(r,\theta,z)} = \frac{1}{V} \int_{r_a}^{r}{\savg{f(r',\theta,z)}r' dr'}
\eeq
Note the identity
\beq
\savg{\gradperp^2 f} = \gradperp^2 \savg{f}
\eeq

For LAPD geometry, the convention is $B_z = - B_p$, so $\bo =
-\zvec$. In normalized variables, $\vE = \bo\times\grad\phi$:
\beqar
V_r = - b_z \frac{1}{r} \pdiff{\phi}{\theta} = \frac{1}{r} \pdiff{\phi}{\theta} \nn \\ 
V_\theta = b_z \pdiff{\phi}{r} = - \pdiff{\phi}{r} \nn \\ 
\eeqar

The equation for the evolution of the azimuthal flows can be obtained
from the vorticity equation (\ref{eqVortA}) by volume averaging.
Applying Gauss theorem and assuming that the boundary conditions on
the internal boundary are such that all surface integrals over the 
internal surface $r=r_a$ vanish, the following expression for the
volume-average of the vorticity is obtained:
\beq\label{eqVortA2}
\vavg{\varpi} = \frac{2\pi L}{V} \savg{N\pdr\phi} 
= -\frac{2\pi L}{V} \savg{N V_\theta} 
\eeq

The ion-ion viscosity term:
\beq\label{eqVortMuA}
\vavg{\gradperp^2\varpi} = \frac{2\pi L}{V} \savg{\pdr\varpi} 
= \frac{2\pi L}{V} \pdr\savg{\varpi} 
\eeq

The advection term can be rewritten as:
\beq
\vavg{\vE\cdot\grad\varpi} = \vavg{\div\left(\varpi\vE\right) - \varpi\div{\vE}}
\eeq
where in straight field the last term vanishes. Applying Gauss
theorem, the full divergence becomes a surface average:
\beq\label{eqVortVeA}
\vavg{\div\left(\varpi\vE\right)}
= \frac{2\pi L}{V} \savg{\varpi V_r} 
\eeq

The fourth term in Eq.~(\ref{eqVortA}) can be written as a full divergence:
\beq
\bvec\times\grad N\cdot\grad\frac{\vE^2}{2} = 
\div\left(\frac{\vE^2}{2}\bvec\times\grad N\right) 
-\frac{\vE^2}{2}\div\left(\bvec\times\grad N\right)=
\div\left(\frac{\gradperp\phi^2}{2}\bvec\times\grad N\right)
\eeq

The volume integral is then transformed info surface average
\beqar\label{eqVortVE2}
\vavg{\bvec\times\grad N\cdot\grad\frac{\vE^2}{2}}
& = &\frac{2\pi L}{V} \savg{\frac{\gradperp\phi^2}{2}\bvec\times\grad N\cdot\hat{r}} \\
\nn
& = &\frac{2\pi L}{V}\frac{1}{r}\savg{\frac{\gradperp\phi^2}{2}\pdth N}
= - \frac{2\pi L}{V}\frac{1}{r}\savg{N\pdth\frac{\gradperp\phi^2}{2}}
\eeqar

Collecting all terms, one obtains the surface-averaged azimuthal momentum evolution equation:
\beq\label{eqmvA}
\pdt\savg{N V_\theta} 
=
 \savg{\varpi V_r} 
+ \frac{1}{r}\savg{N\pdth\frac{\gradperp\phi^2}{2}}
- \nuin\savg{N V_\theta} 
- \mu\pdr\savg{\varpi} 
\eeq
%

%%%%%%%%%%%%%%%%%%%%%%%%%%%%%%%%%%%%%%%%%%%%%%%%%%%%%%%%%%%%%%%%%%%%%%%%%%%

%%%%%%%%%%%%%%%%%%%%%%%%%%%%%%%%%%%%%%%%%%%%%%%%%%%%%%%%%%%%%%%%%%%%%%%%%%%

\section{Ion viscosity}\label{secAppendixVisc}

Including the ion-ion viscosity effects results in an additional term
in the vorticity equation~(\ref{eqVort}). Viscous force $\div\Pi$
(where $\Pi$ is the stress tensor $\vvec\vvec-v^2/3{\bf\hat{I}}$) in the ion motion
equation induces a perpendicular drift
$\vvec_{i\perp\mu}=\bvec\times\left(\div\Pi\right)$ to the lowest
order, which translates into an additional perpendicular current
component $\jvec_{\perp,\mu} = en\bvec\times\left(\div\Pi\right)$. To
simplify the final form of the viscous term, it is assumed that
density perturbations are small and the equilibrium gradients are
negligible compared to the perturbed quantities. Then the extra term
due to viscosity in the vorticity equation can be written as
\beq
\div j_{\perp,\mu} = en\div\left\{\bvec\times\left(\div\Pi\right)\right\}
= -en\bvec\cdot\left\{\grad\times\left(\div\Pi\right)\right\}
\eeq
Using Braginskii expressions for the stress
tensor~\cite{Braginskii1965} with perpendicular ion velocity
$\vvec_{i\perp}=\vE=\bvec\times\grad\phi$ to the lowest order, it can
be shown that the extra term in the vorticity equation~(\ref{eqVort}) is
$\mu_i\gradperp^2\varpi$.

Depending on the ion magnetization, one should choose either the
magnetized or unmagnetized viscosity expression for the coefficient
$\mu_i$: $\mu_i=\eta_1^i=0.3nT_i/\wci^2\tau_i$ for $\wci\tau_i\gg 1$
or $\mu_i=\eta_0^i=0.96nT_i\tau_i$ for $\wci\tau_i\ll 1$. The estimate
of the ion magnetization parameter $\wci\tau_i$ for typical LAPD
values (${\rm He}^4$, $B_0=400~\G$, $n_i\sim\enum{3}{18}~\mn$,
$T_i\sim 1~\eV$) is close to unity.  The ion temperature in LAPD is
not directly measured; the estimate from the electron-ion energy
exchange and the parallel losses balance indicates that $T_i$ is in the
range $0.1-1~\eV$. Preliminary BOUT simulations with ion viscosity are consistent
with the experimental measurements for low ion temperatures, $T_i\sim
0.1~\eV$, which corresponds to unmagnetized ion regime.

%%%%%%%%%%%%%%%%%%%%%%%%%%%%%%%%%%%%%%%%%%%%%%%%%%%%%%%%%%%%%%%%%%%%%%%%%%%

\section{Discretization in the parallel coordinate}\label{secAppendixGrid}

Consider the local drift mode dispersion relation for the simplest
case, without electron inertia and electromagnetic terms, when it
becomes a quadratic equation, as given in elementary plasma textbooks~\cite{Chen1984}:
\beq
(\omega - 1) 
i \sigma_{||}
+ 
\omega^2 
= 0,
\label{e:disp_rel_simple}
\eeq
where 
\beq
\sigma_{||} 
= 
\left( \frac{k_{||}}{k_{\perp}} \right)^2 
\frac{\Omega_{ci} \Omega_{ce}}
{\nu_{ei}\omega_*},
\eeq
\beq
\omega_*= k_{\perp} v_{pe} = \frac{k_\perp}{L_n}\frac{T_{e0}}{m_i\Omega_{ci}}
\eeq
and $\omega$ is normalized to $\omega_*$.

Now consider the effect of finite-difference discretization on
Eq.~(\ref{e:disp_rel_simple}). For simplicity assume no radial
structure so that the radial part of the solution can be
dropped. There are just two coordinates then - the drift wave
propagation direction $y$ and the parallel direction $z$.

Now let's focus on the parallel discretization. Parallel derivatives
that are represented by $i k_{||}$ in the Fourier form will become
something different in the discretized equation, depending on the type
of discretization. For example, applying the 2$^{nd}$ central
difference for a single mode $\exp(i k_{||} z)$ one obtains
\beq
\frac{df}{dz} 
\rightarrow 
\frac{\exp(i k_{||} z_{j+1}) - \exp(i k_{||} z_{j-1})} {2 h}
=
\frac{i\sin(k_{||} h)}{h}
\eeq
Here $h$ is parallel grid spacing, $z_j$=$j h$. 

One can notice that at large wavenumbers, $k \sim \pi/h$, the
finite-difference representation is very poor. In this example, from
Eq.~{\ref{e:disp_rel_simple}}, for ${\sigma_{||}} \gg$1 the
growth rate scales as $\gamma \propto 1/k_{||}^2$, i.e., large
$k_{||}$ should stabilize the modes. Conversely, in the discretized
dispersion relation it will become $\gamma \propto 1/(\sin(k_\para h))^2$
which would become singular at the Nyquist wavenumber $k_\para = \pi/h$,
which can be manifested in unphysical behavior of such modes. The
possibility of unphysical behavior of high-k modes caused by spatial
discretization, in particular the ``red-black'' numerical instability,
is well-known in the CFD community, and historically the main remedy
was using staggered grid \cite{Patankar1980}. A more recent popular
method is discretization on collocated grids adding a dissipative
biharmonic (i.e. 4$^{th}$ derivative) term to suppress the
``red-black'' instability (e.g., the Rhie-Chow interpolation).

Consider the effects of staggered grid for the discretized drift mode
dispersion relation. Assume that $N_i$, $\phi$, $\varpi$ are specified
on one grid, and $j_{||}$, $V_{||e}$ are specified on another grid
shifted by $h$/2. Then, $k_{||}^2$ in the dispersion relation
(\ref{e:disp_rel_simple}) can be tracked down to the derivatives
\beq
\pdiff{j_{||}}{z} \rightarrow (j_{||,j}-j_{||,j-1})/h =
\exp(-i k h /2) \frac{i \sin(k h /2)}{h/2}
\label{e:leftfd}
\eeq
and
\beq
\pdiff{\phi_{||}}{z} \rightarrow (\phi_{j+1}-\phi_{j})/h =
\exp(i k h /2) \frac{i \sin(k h /2)}{h/2},
\label{e:rightfd}
\eeq
which combine to produce
\beq
k_{||}^2 \rightarrow - 
\left(
\frac{\sin(k h /2)}{h/2}
\right)^2
\label{e:fixed_2nd_deriv}
\eeq
One can note that Eq.~(\ref{e:fixed_2nd_deriv}) does not become zero
for any mode supported by the grid, $-\pi/h \le k \le \pi/h$, which is
an important improvement.

Since the staggered grid approach is more cumbersome for
implementation, in particular for parallel computation, it is
desirable to stay with collocated grids, if possible. In this example
one can come up with discretization
Eq.~(\ref{e:leftfd}-\ref{e:rightfd}) on collocated grids by combining
right-sided and left-sided 1$^{st}$ order discretization. This method,
which we call ``quasi-staggered grid'', has been successfully applied
in BOUT to eliminate spurious modes due to parallel discretization.

%%%%%%%%%%%%%%%%%%%%%%%%%%%%%%%%%%%%%%%%%%%%%%%%%%%%%%%%%%%%%%%%%%%%%%%%%%%

%\bibliographystyle{phaip}
%\bibliography{bout_lapd_refs}

\begin{thebibliography}{10}


\bibitem{schekochihin2006}
A.~Schekochihin and S.~Cowley,
\newblock Phys. Plasmas {\bf 13}, 056501 (2006).

\bibitem{quataert2002}
E.~Quataert, W.~Dorland, and G.~W. Hammett,
\newblock ApJ {\bf 577}, 524 (2002).

\bibitem{hasegawa2004}
H.~Hasegawa, M.~Fujimoto, T.~D.Phan, H.~Reme, A.~Balogh, M.~W. Dunlop, C.~Hashimoto, and R.~TanDokoro,
\newblock Nature {\bf 430}, 755 (2004).

\bibitem{tynan2009}
G.~Tynan, A.~Fujisawa, and G.~McKee,
\newblock Plasma Phys. Control. Fusion {\bf 51}, 113001 (2009).

\bibitem{connor1995}
J.~Connor,
\newblock Plasma Phys. Control. Fusion {\bf 37}, A119 (1995).

\bibitem{horton1999}
W.~Horton,
\newblock Rev. Mod. Phys. {\bf 71}, 735 (1999).

\bibitem{zweben2007}
S.~J. Zweben, J.~A. Boedo, O.~Grulke, C.~Hidalgo, B.~LaBombard, R.~J. Maqueda, P.~Scarin, and J.~L. Terry,
\newblock Plasma Phys. Control. Fusion {\bf 49}, S1 (2007).

\bibitem{Kadomtsev1965}
B.~B. Kadomtsev,
\newblock {\em Plasma Turbulence},
\newblock Academic Press, London, 1965.

\bibitem{Stenflo1970}
L.~Stenflo,
\newblock J. Plasma Physics {\bf 4}, 585 (1970).

\bibitem{cowley1991}
S.~C. Cowley, R.~M. Kulsrud, and R.~Sudan,
\newblock Phys. Fluids {\bf B 3}, 2767 (1991).

\bibitem{dorland2000}
W.~Dorland, F.~Jenko, M.~Kotschenreuther, and B.~Rogers,
\newblock Phys. Rev. Lett. {\bf 85}, 5579 (2000).

\bibitem{Iroshnikov1963}
P.~S. Iroshnikov,
\newblock AZh {\bf 40}, 742 (1963).

\bibitem{Kraichnan1965}
R.~H. Kraichnan,
\newblock Phys. Fluids {\bf 8}, 1385 (1965).

\bibitem{Itoh1988}
S.-I. Itoh and K.~Itoh,
\newblock Phys. Rev. Lett. {\bf 60}, 2276 (1988).

\bibitem{Shaing1989}
K.~C. Shaing and E.~C. Crume,
\newblock Phys. Rev. Lett. {\bf 63}, 2369 (1989).

\bibitem{Biglari1990}
H.~Biglari, P.~H. Diamond, and P.~W. Terry,
\newblock Phys. Fluids {\bf B 2} (1990).

\bibitem{Greenwald2010}
M.~Greenwald,
\newblock Phys. Plasmas {\bf 17}, 058101 (2010).

\bibitem{Terry2008}
P.~W. Terry, M.~Greenwald, J.-N.~Leboeuf, G.~R. McKee, D.~R. Mikkelsen, W.~M. Nevins, D.~E. Newman, and D.~P. Stotler,
\newblock Phys. Plasmas {\bf 15}, 062503 (2008).

\bibitem{Gekelman1991}
W.~Gekelman, H.~Pfister, Z.~Lucky, J.~Bamber, D.~Leneman, and J.~Maggs,
\newblock Rev. Sci. Inst. {\bf 62}, 2875 (1991).

\bibitem{Tynan2004}
G.~R. Tynan, M.~J. Burin, C.~Holland, G.~Antar, and P.~H. Diamond,
\newblock Plasma Phys. Control. Fusion {\bf 46}, A373 (2004).

\bibitem{Franck2002}
C.~M. Franck, O.~Grulke, and T.~Klinger,
\newblock Phys. Plasmas {\bf 9}, 3254 (2002).

\bibitem{Shinohara1995}
S.~Shinohara, Y.~Miyauchi, and Y.~Kawai,
\newblock Plasma Phys. Control. Fusion {\bf 37}, 1015 (1995).

\bibitem{Lynn2009}
A.~G. Lynn, M.~Gilmore, C.~Watts, J.~Herrea, R.~Kelly, S.~Will, S.~Xie, L.~Yan, and Y.~Zhang,
\newblock Rev. Sci. Inst. {\bf 80}, 103501 (2009).

\bibitem{Pierre1987}
T.~Pierre, G.~Leclert, and F.~Braun,
\newblock Rev. Sci. Inst. {\bf 58}, 6 (1987).

\bibitem{kasuya2007}
N.~Kasuya, M.~Yagi, M.~Azumi, K.~Itoh, and S.-I. Itoh,
\newblock J. Phys. Soc. Japan {\bf 76}, 044501 (2007).

\bibitem{Holland2007}
C.~Holland, G.~R. Tynan, J.~H. Yu, A.~James, D.~Nishijima, M.~Shimada, and N.~Taheri,
\newblock Plasma Phys. Control. Fusion {\bf 49}, A109 (2007).

\bibitem{Naulin2008}
V.~Naulin, T.~Windisch, and O.~Grulke,
\newblock Phys. Plasmas {\bf 4}, 012307 (2008).

\bibitem{Kervalishivili2008}
G.~N. Kervalishvili, R.~Kleiber, R.~Schneider, B.~D. Scott, O.~Grulke, and T.~Windisch,
\newblock Contrib. Plasma Phys. {\bf 48}, 32 (2008).

\bibitem{Windisch2008}
T.~Windisch, O.~Grulke, R.~Schneider, and G.~N. Kervalishvili,
\newblock Contrib. Plasma Phys. {\bf 48}, 58 (2008).

\bibitem{Rogers2010}
B.~N. Rogers and P.~Ricci,
\newblock Phys. Rev. Lett. {\bf 104}, 225002 (2010).

\bibitem{Xu1998}
X.~Q. Xu and R.~H. Cohen,
\newblock Contrib. Plasma Phys. {\bf 36}, 158 (1998).

\bibitem{Umansky2009}
M.~Umansky, X.~Xu, B.~Dudson, L.~LoDestro, and J.~Myra,
\newblock Contrib. Plasma Phys. {\bf 180}, 887 (2009).

\bibitem{Popovich2010a}
P.~Popovich, M.~Umansky, T.~A. Carter, and B.~Friedman,
\newblock Phys. Plasmas {\bf 17}, 102107 (2010).

\bibitem{carter2009}
T.~A. Carter and J.~E. Maggs,
\newblock Phys. Plasmas {\bf 16}, 012304 (2009).

\bibitem{maggs2007}
J.~E. {Maggs}, T.~A. {Carter}, and R.~J. {Taylor},
\newblock Phys. Plasmas {\bf 14}, 052507 (2007).

\bibitem{Braginskii1965}
S.~I. Braginskii,
\newblock T{ransport processes in a plasma},
\newblock in {\em Reviews of Plasma Physics}, edited by M.~A. Leontovich,
  volume~1, pages 205--311, Consultants Bureau, New York, 1965.

\bibitem{Vincena2001}
S.~Vincena, W.~Gekelman, and J.~E. Maggs,
\newblock Phys. Plasmas {\bf 8}, 3884 (2001).

\bibitem{maggs2003}
J.~E. Maggs and G.~J. Morales,
\newblock Phys. Plasmas {\bf 10}, 2267 (2003).

\bibitem{penano1997}
J.~R.~Pe\~{n}ano, G.~J. Morales, and J.~E. Maggs,
\newblock Phys. Plasmas {\bf 4}, 555 (1997).

\bibitem{ono1975}
M.~Ono and R.~Kulsrud,
\newblock Phys. Fluids {\bf 18}, 1287 (1975).

\bibitem{Dudson2009}
B.~Dudson,
\newblock private communication, 2009.

\bibitem{pace2008}
D.~Pace, M.~Shi, J.~Maggs, G.~Morales, and T.~Carter,
\newblock Phys. Rev. Lett. {\bf 101}, 085001 (2008).

\bibitem{Carter2006}
T.~A. Carter,
\newblock Phys. Plasmas {\bf 13}, 010701 (2006).

\bibitem{Tynan2006}
G.~R. Tynan, C.~Holland, J.~H. Yu, A.~James, D.~Nishijima, M.~Shimada, and N.~Taheri,
\newblock Plasma Phys. Control. Fusion {\bf 48}, S51 (2006).

\bibitem{Tsai1970}
S.-T. Tsai, F.~W. Perkins, and T.~H. Stix,
\newblock Phys. Fluids {\bf 13}, 2108 (1970).

\bibitem{ricci2006}
P.~Ricci, B.~N. Rogers, W.~Dorland, and M.~Barnes,
\newblock Phys. Plasmas {\bf 13}, 062102 (2006).

\bibitem{Berk1991}
H.~L. Berk, D.~D. Ryutov, and Y.~Tsidulko,
\newblock Phys. Fluids {\bf B 3}, 1346 (1991).

\bibitem{Boedo2003}
J.~A. Boedo, D.~L. Rudakov, R.~A. Moyer, G.~R. McKee, R.~J. Colchin, M.~J. Schaffer, P.~G. Stangeby, W.~P. West, S.~L. Allen, T.~E. Evans, R.~J. Fonck, E.~M. Hollmann, S.~Krasheninnikov, A.~W. Leonard, W.~Nevins, M.~A. Mahdavi, G.~D. Porter, G.~R. Tynan, D.~G. Whyte, and X.~Xu",
\newblock Phys. Plasmas {\bf 10}, 1670 (2003).

\bibitem{Boedo2001}
J.~A. Boedo, D.~Rudakov, R.~Moyer, S.~Krasheninnikov, D.~Whyte, G.~McKee, G.~Tynan, M.~Schaffer, P.~Stangeby, P.~West, S.~Allen, T.~Evans, R.~Fonck, E.~Hollmann, A.~Leonard, A.~Mahdavi, G.~Porter, M.~Tillack, and G.~Antar,
\newblock Phys. Plasmas {\bf 8}, 4826 (2001).

\bibitem{Chen1984}
F.~F. Chen,
\newblock {\em {I}ntroduction to {P}lasma {P}hysics and {C}ontrolled {F}usion},
\newblock Plenum Press, New York, 1984.

\bibitem{Patankar1980}
S.~V. Patankar,
\newblock {\em Numerical Heat Transfer and Fluid Flow},
\newblock Hemisphere Publishing Corporation, New York, 1980.





\end{thebibliography}

%
%
\end{document}